\documentclass[sigconf,nonacm]{acmart}
\usepackage{array}
\usepackage{graphicx}
\usepackage{subfig}
\usepackage{caption}
\usepackage{balance}
\usepackage{float}
\AtBeginDocument{%
  }

\begin{document}

\title[Raising the Bar(ometer): Identifying a User's Stair and Lift Usage Through Wearable Sensor Data Analysis]{
Raising the Bar(ometer): Identifying a User's Stair and Lift Usage \\Through Wearable Sensor Data Analysis}

\author{Hrishikesh Balkrishna Karande}
\orcid{0009-0005-8223-6395}
\authornote{The first five authors made equal contributions to this research.}

\author{Ravikiran Arasur Thippeswamy Shivalingappa}
\orcid{0009-0006-0565-4849}
\authornotemark[1]
\affiliation{%
  \institution{University of Siegen}
  \city{Siegen}
  \country{Germany}
}

\author{Abdelhafid Nassim Yaici}
\orcid{0009-0004-1319-4606}
\authornotemark[1]

\author{Iman Haghbin}
\orcid{0009-0003-3788-2326}
\authornotemark[1]

\author{Niravkumar Bavadiya}
\orcid{0009-0009-7900-3047}
\authornotemark[1]
\affiliation{%
  \institution{University of Siegen}
  \city{Siegen}
  \country{Germany}
}

\author{Robin Burchard}
\orcid{0000-0002-4130-5287}

\author{Kristof Van Laerhoven}
\orcid{0000-0001-5296-5347}
\affiliation{%
  \institution{University of Siegen}
  \city{Siegen}
  \country{Germany}
}

\renewcommand{\shortauthors}{Hrishikesh Karande, Ravikiran Arasur, Nassim Yaici, Iman Haghbin, Nirav Bavadiya et al.}

\begin{abstract}
Many users are confronted multiple times daily with the choice of whether to take the stairs or the elevator. 
Whereas taking the stairs could be beneficial for cardiovascular health and wellness, taking the elevator might be more convenient but it also consumes energy. By precisely tracking and boosting users' stairs and elevator usage through their wearable, users might gain health insights and motivation, encouraging a healthy lifestyle and lowering the risk of sedentary-related health problems. 
This research describes a new exploratory dataset, to examine the patterns and behaviors related to using stairs and lifts. We collected data from 20 participants while climbing and descending stairs and taking a lift in a variety of scenarios. The aim is to provide insights and demonstrate the practicality of using wearable sensor data for such a scenario. 
Our collected dataset was used to train and test a Random Forest machine learning model, and the results show that our method is highly accurate at classifying stair and lift operations with an accuracy of 87.61\% and a multi-class weighted F1-score of 87.56\% over 8-second time windows. Furthermore, we investigate the effect of various types of sensors and data attributes on the model's performance. Our findings show that combining inertial and pressure sensors yields a viable solution for real-time activity detection.
\end{abstract}



\maketitle

\section{Introduction}
The proliferation of wearable technology in recent years has significantly transformed how we monitor and understand human physical activities. Wearable devices, equipped with a variety of sensors, have emerged as pivotal tools in numerous fields including health monitoring, fitness tracking, and medical diagnostics~\cite{mattmann2007recognizing, patel2012review}.
Activity recognition, a key application of wearable technology, involves the identification and classification of various physical actions performed by an individual, such as walking, running, sitting, or in our case more complex activities like climbing stairs or using an elevator~\cite{randell2000context, attal2015physical}.

Differentiating between activities such as stair climbing and lift use is especially important since each activity places different physiological and bio-mechanical demands on the body~\cite{nickblackmer2023}. Stair climbing is a physically challenging activity that can yield useful information on cardiovascular health, lower body strength, and overall mobility. It is also connected with a high caloric expenditure, making it an important exercise to track in fitness and weight control settings~\cite{atlas,captaincalc2020}. In contrast, using an elevator indicates stationary behavior, emphasizing times of immobility that are critical to understanding overall activity patterns and such behavior \cite{rosalynn2024}.

The use of pressure sensors in wearable electronics provides a precise and effective approach for capturing the finer details of these actions. Pressure sensors detect differences in pressure at different positions or in various situations, providing extensive information on an individual's gait and posture~\cite{kim2020recent}. For example, the pattern of pressure changes while ascending stairs differs significantly from the pressure change while standing still in an elevator. Using these sensors, we can improve the accuracy of activity recognition, allowing us to distinguish between the energy demands of stair climbing and the passive nature of the elevator use.

The ability to accurately recognize and differentiate various activities offers profound implications across multiple domains. In personalized healthcare, it enables tailored interventions and monitoring, enhancing treatment plans and preventive care \cite{wu2019wearable}. Additionally, In the context of elderly care, detecting the specific activity of stair climbing versus elevator use becomes critical in ensuring safety. Many older adults are at a higher risk of falls when ascending or descending stairs, and timely detection of stair use can help caregivers monitor and prevent potential accidents or offer immediate assistance when needed\cite{delahoz2014survey}. In this way, monitoring specific activities can contribute to safer and more secure living environments for vulnerable populations. In the fitness industry, these technologies allow for more customized exercise plans and real-time feedback on performance, aiding in achieving personal fitness goals \cite{thompson2019worldwide}. Furthermore, In workplace ergonomics and occupational health, recognizing whether employees are using stairs or elevators can assist in designing more efficient movement patterns within the workplace. Employers can encourage healthier habits such as stair climbing, which can reduce sedentary behavior, improve cardiovascular health, and enhance overall well-being among workers\cite{kritzler2015wearable}. Additionally, analyzing activity patterns can help assess whether employees are adhering to ergonomic safety protocols, potentially reducing the risk of workplace injuries associated with improper movements or excessive inactivity.

Our dataset\footnote{Both the code and data (raw and resampled form) for the project are publicly accessible on GitHub: \url{https://github.com/iiMox/project_work_stairs_lift_detection.git}} and this paper's preliminary experiments offer several contributions:
\begin{itemize}
    \item Introduction of a stairs and lift taking dataset: We present a dataset specifically dedicated to distinguishing between lift usage and stair climbing with multiple sensors and a high variability between participants.
    \item Characterization of barometric pressure sensor data: We provide a proof of concept for the application of pressure sensor data in activity recognition.
    \item Examination of Window Size for Time-Series Data: We investigate the impact of various window sizes on the pre-processing of time-series data.
    \item Feature Importance Analysis: We analyze the importance of different features, highlighting the critical role of pressure measurements in accurately classifying activities.
\end{itemize}

\section{Related Technologies}
\subsection{Sensors}
The field of activity recognition using wearable technology has seen significant advancements due to the integration of various sensor technologies and data analysis techniques~\cite{zhang2022deep}.
In our study, we utilized accelerometer and barometer sensors to capture the necessary data for activity recognition. The accelerometer measures the acceleration forces acting on the device, providing detailed information about movement and orientation. The barometer, on the other hand, measures atmospheric pressure, which can be useful in detecting changes in elevation, such as when a person uses stairs or an elevator~\cite{ramasamy2018recent}.

\subsection{State of the art and limitations}

The development and validation of activity recognition systems have heavily relied on various publicly available datasets. These datasets contain sensor data collected from wearable devices, which were placed on participant's bodies and recorded during different physical activities. The datasets are crucial for training and evaluating machine learning models. This section reviews some of the most commonly used datasets in the field that include stair usage, highlighting their contributions.

Among the prominent datasets, the UCI HAR~\cite{anguita2013public} dataset includes activities such as "walking upstairs" and "walking downstairs" performed by participants wearing smartphones on their waists. Similarly, the PAMAP2~\cite{6246152} dataset captures a range of activities including "ascending stairs" and "descending stairs" using multiple sensors placed on participants' bodies. The MHealth dataset~\cite{banos2014mhealthdroid} also includes "climbing stairs" along with other physical exercises. Table~\ref{tab:commands} lists some of the public activity recognition datasets that contain ``stairs'' as one of their activities. The table also includes information about the recorded sensors and the overall size of the datasets mentioned. 

None of the datasets we examined that included stair usage involved the activity of taking lifts. Furthermore, to our knowledge, there are no datasets that specifically involve the activity of using lifts. There are also devices like ``Fitbit'' that can detect if you are taking the stairs~\cite{thetechylife2024}, using sensors such as altimeters to measure elevation changes. However, these devices do not focus on differentiating between the usage of lifts and stairs specifically, and user reports~\cite{ivanjovin2019,tomnielson2023} indicate that they cannot reliably detect the difference between lifts and stairs.

In addition to these datasets, other related works pursue the same goal of promoting healthy living through wearable technology. For instance, Neves et al.~\cite{neves2008application} explore the use of wireless sensor networks to monitor health metrics and encourage healthier behaviors. Similarly, Ohtaki et al.~\cite{ohtaki2005automatic} developed a system that automatically classifies ambulatory movements and estimates energy consumption using accelerometers and a barometer. Although these studies do not focus on differentiating between lift and stair usage, they contribute to the broader objective of using wearable devices and sensor technologies to enhance health by monitoring physical activity and energy expenditure, aligning closely with our study’s focus on leveraging sensor technology for health promotion.

Our study, therefore, aims to fill this gap by providing extensive sensor data from participants specifically using lifts and stairs.

Our focus is on capturing the use of lifts and stairs, and not just having incidental usage. Our study, therefore, aims to fill this gap by providing extensive sensor data from a larger cohort of participants specifically using lifts and stairs. This level of detail and focus on lift versus stair usage is not comprehensively covered by existing datasets, making our publicly available dataset the first with this focus.

\begin{table*}
  \label{tab:commands}

\begin{tabular}{|m{2.7cm}|m{4cm}|m{.9cm}|m{.9cm}|m{.9cm}|m{3.5cm}|m{2.5cm}|}
    \hline
    \textbf{Dataset} & \textbf{Activities} & \textbf{Stairs} & \textbf{Lift} & \textbf{Users} & \textbf{Sensors used} & \textbf{Size of dataset} \\
    \hline
    UCI HAR (Human Activity Recognition Using Smartphones) \cite{anguita2013public} & Walking, Walking Upstairs, Walking Downstairs, Sitting, Standing, Laying & Yes & No & 30 & Accelerometer, Gyroscope & 10,299 Observations (Each 2.56 seconds)\\
    \hline
    PAMAP2 (Physical Activity Monitoring) \cite{6246152} & 18 total - basic, household, and exercise activities & Yes & No & 9 & Accelerometer, Gyroscope, Magnetometer, Heart Rate Monitoring & About 10 hours (IMU data: 100 Hz Heart rate data: 9 Hz)\\
    \hline
    WISDM \cite{misc_wisdm_smartphone_and_smartwatch_activity_and_biometrics_dataset__507,weiss2019wisdm} & 18 total - ambulation-related activities, hand-based activities of daily living, and various eating activities & Yes & No & 51 & Accelerometer, Gyroscope & 2754 minutes\\
    \hline
    MHealth \cite{banos2014mhealthdroid} & 12 total - Basic, Locomotion, and exercise activities & Yes & No & 10 & Accelerometer, Gyroscope, Magnetometer, Electrocardiogram (heart monitoring) & Not specified\\
    \hline
    Ours & Walking Upstairs, Walking Downstairs, Lift up, Lift down, Null & Yes & Yes & 20 & Accelerometer, Barometer & 525.02 minutes (50~Hz)\\
    \hline
\end{tabular}

    \caption{State-of-the-art datasets that are related to activity recognition and contain going up or down the stairs as an activity. The datasets differ in the contained activities, numbers of participants, and sensors used, as well as in their sizes. Our newly created dataset is the only available dataset that includes taking the lift as an activity and uses a barometric pressure sensor. }
\end{table*}

\section{Data Collection \& Analysis}
A total of 20 participants were involved in the data collection process. These participants varied in age (26.0 ± 10.75), gender (10 male and 10 female), and physical fitness levels to ensure a diverse dataset that reflects different human movement patterns. Prior to participation, each individual signed a consent form, acknowledging their voluntary involvement and understanding of the experiment’s procedures and objectives. Additionally, we submitted a detailed proposal to The Council for Ethics in Research (Ethics Council) of the University of Siegen, which reviewed and approved the study, allowing us to proceed with the experiment. The data was gathered using the Bangle.js 2 smartwatch, which is equipped with both accelerometer and barometer sensors. The accelerometer measures the acceleration forces acting on the device in three-dimensional space, while the barometer provides data on atmospheric pressure, which can be used to infer altitude changes.

\subsection{Data-collection Procedure}
The experiment was conducted in a university building that comprises a total of eight floors that can be switched between by either flights of stairs or a set of elevators. This environment was chosen due to its accessibility and the availability of both stairs and elevators. To ensure variability and reduce systematic bias, we employed randomization in our experiment in several ways:

\begin{itemize}
\item The next floor for each participant to visit was determined by a random number generator, ensuring an unpredictable sequence of floor visits. The floor from which the experiment would be started was also determined by the random number generator. Since the university building that we used has floors numbered from 2 (first accessible floor) to 8 (last accessible floor), our random number generator had been set to produce the lowest possible random number 2 and the highest possible random number 8.      
\item The choice between using the elevator or stairs for each transition was decided by a coin toss, ensuring an impartial probability of selecting either mode of movement. 
\item The participants were advised to act normal and try different variations of activities that they would ideally do while performing the movement of taking the stairs or taking the Lifts. Some participants chose to carry a handbag or a set of books with them, while others used their phones while taking the lifts or stairs. The participants were encouraged to do the experiment at their own pace and with their own choice of effort level. These variations of the different hand movements that the participants did while participating in the experiment were recorded using the accelerometer.
\end{itemize}

Each participant was instructed to wear the Bangle.js 2 smartwatch throughout the experiment. They were guided to move between floors based on the outcomes of the random number generator and the coin toss. The smartwatch continuously recorded accelerometer and barometer data during these activities at a frequency of 50 Hz.
For each participant, the experiment was conducted over a period of approximately 30 minutes. Participants were given adequate breaks between movements to ensure they were not fatigued, which could affect the naturalness of their movements. The data collection sessions were scheduled to avoid peak hours in the building, minimizing external disruptions and ensuring safety.

Each participant was accompanied by four researchers, each one of these four were responsible for one of these tasks:
\begin{enumerate}
    \item Video recording the participant's complete physique for the entire duration of the experiment.
    \item Annotating the changes in the events that are taking place, classifying them into the 5 classes, and also noting down the time when this change took place.
    \item Generating a random number to decide which floor to go to next.
    \item Tossing the coin to decide whether to take the stairs or the lift to this floor.
\end{enumerate}

During the data annotation phase, using the video recordings and based on the context data from the annotation data files, class labels were manually added to the sensor data CSV (Comma-Separated Values) files as ground truth for the experiment.

\subsection{Activity Classes}
The dataset was categorized into five distinct classes:
\begin{itemize}
    \item \textbf{Null}: Periods in which none of the target activities was performed, includes idleness (such as sitting, standing still, walking, or pacing).
    \item \textbf{Lift Up}: Movement of participants using the elevator to ascend floors.
    \item \textbf{Lift Down}: Movement of participants using the elevator to descend floors.
    \item \textbf{Stairs Up}: Movement of participants using the stairs to ascend floors.
    \item \textbf{Stairs Down}: Movement of participants using the stairs to descend floors.
\end{itemize}
We gathered a total of 525 minutes (8:45h) of data from the participants. The distribution of this data among the five classes can be seen in Figure \ref{fig:data_distribution}. The exact duration of each class in minutes is given in appendix D table \ref{duration_table}.
\begin{figure}
    \centering
    \includegraphics[width=\linewidth]{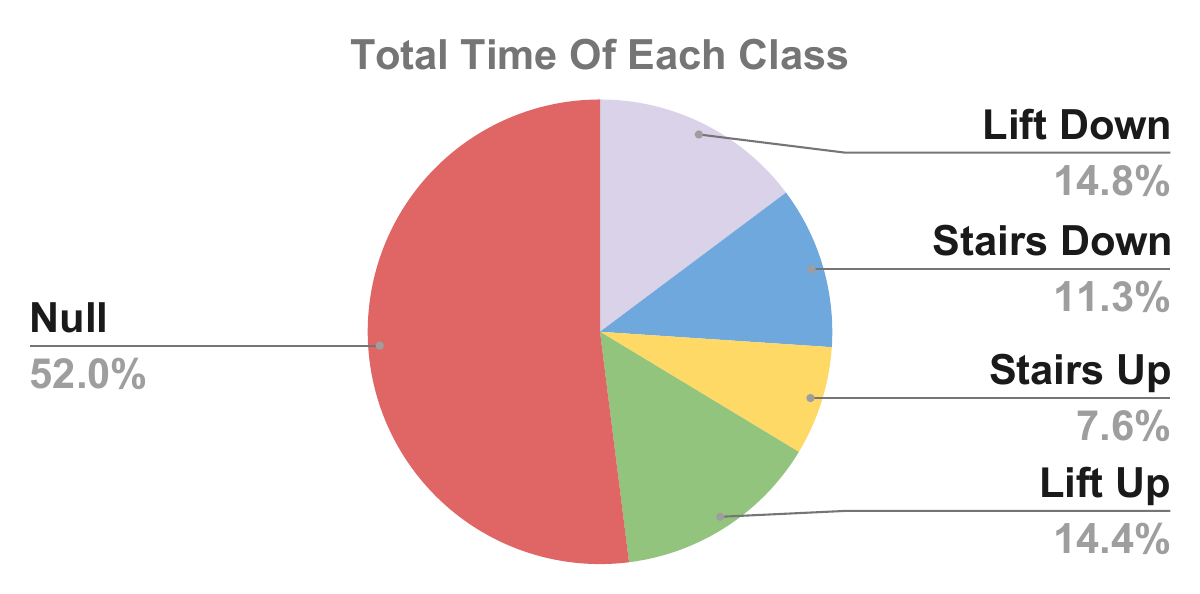}
    
    \caption{The distribution of the activity classes contained in the recorded dataset. The Null class makes up 52\,\%, while the rest of the data consists of the four target classes (Lift Up, Lift Down, Stairs Up, Stairs Down).} 
    \label{fig:data_distribution}
\end{figure}

\subsection{Analysis}
\subsubsection{Sensor Data: }
The sensor data contains detailed information about the movement and pressure measurements taken at various time intervals. This data is evaluated to better comprehend movement patterns, acceleration variations, and possible activities or behaviors.
\begin{itemize}
    \item Time: The timestamp indicating the time at which the data was recorded.
    \item Timestamp: A numerical representation of time, in milliseconds.
    \item X, Y, Z: Accelerometer data representing the acceleration along the X, Y, and Z axes, respectively.
    \item Magnitude: Magnitude of the acceleration vector.
    \item Pressure: Pressure data.
    \item Label: Ground truth label for the data. One of: Stairs Up, Stairs down, Lift up, Lift down \& Null.
\end{itemize}

\begin{figure}
    \centering
    \includegraphics[width=\linewidth,trim={1.2cm 1.2cm 1.2cm 1.2cm},clip]{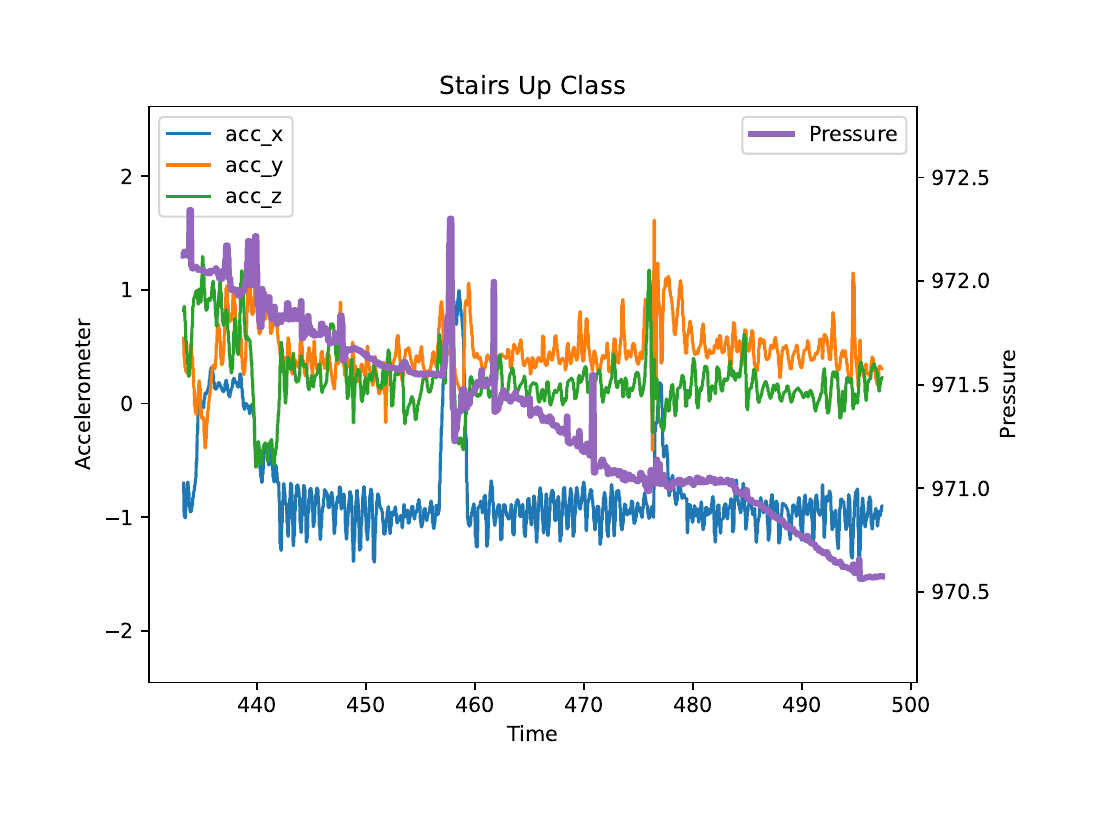}
    
    \caption{Example accelerometer and barometer data for the 'Stairs Up' activity class. The accelerometer shows the typical patterns of walking. Meanwhile, the pressure steadily decreases, with short plateaus occurring on the stairway's landings. Additionally, the pressure sensor shows some spikes which are likely caused by the movement of the arm while walking.} 
    \label{fig:sensor_data-SU}
\end{figure}
\begin{figure}
    \centering
    \includegraphics[width=\linewidth,trim={1.2cm 1.2cm 1.2cm 1.2cm},clip]{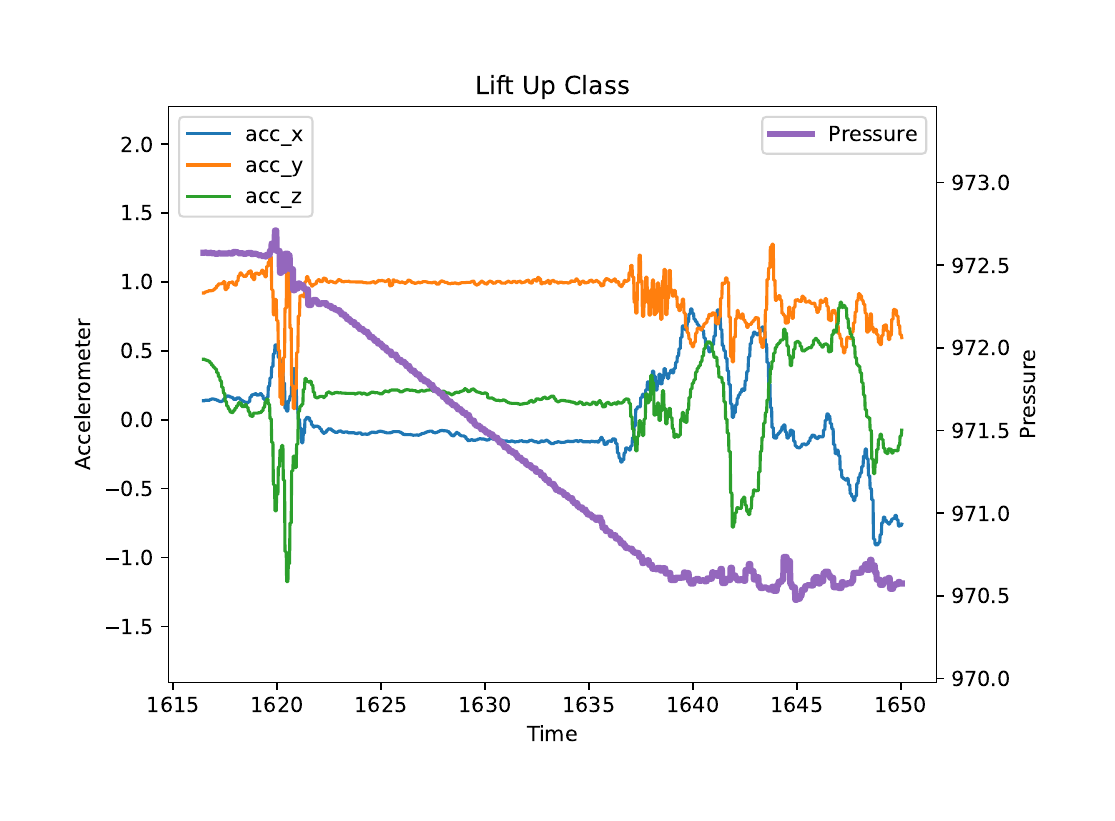}
    
    \caption{Example accelerometer and barometer data for the 'Lift Up' activity class. While the lift is moving, the participant mostly stood still. We observe the pressure decreasing smoothly.} 
    \label{fig:sensor_data-LU}
\end{figure}
\begin{figure}
    \centering
    \includegraphics[width=\linewidth,trim={1.2cm 1.2cm 1.2cm 1.2cm},clip]{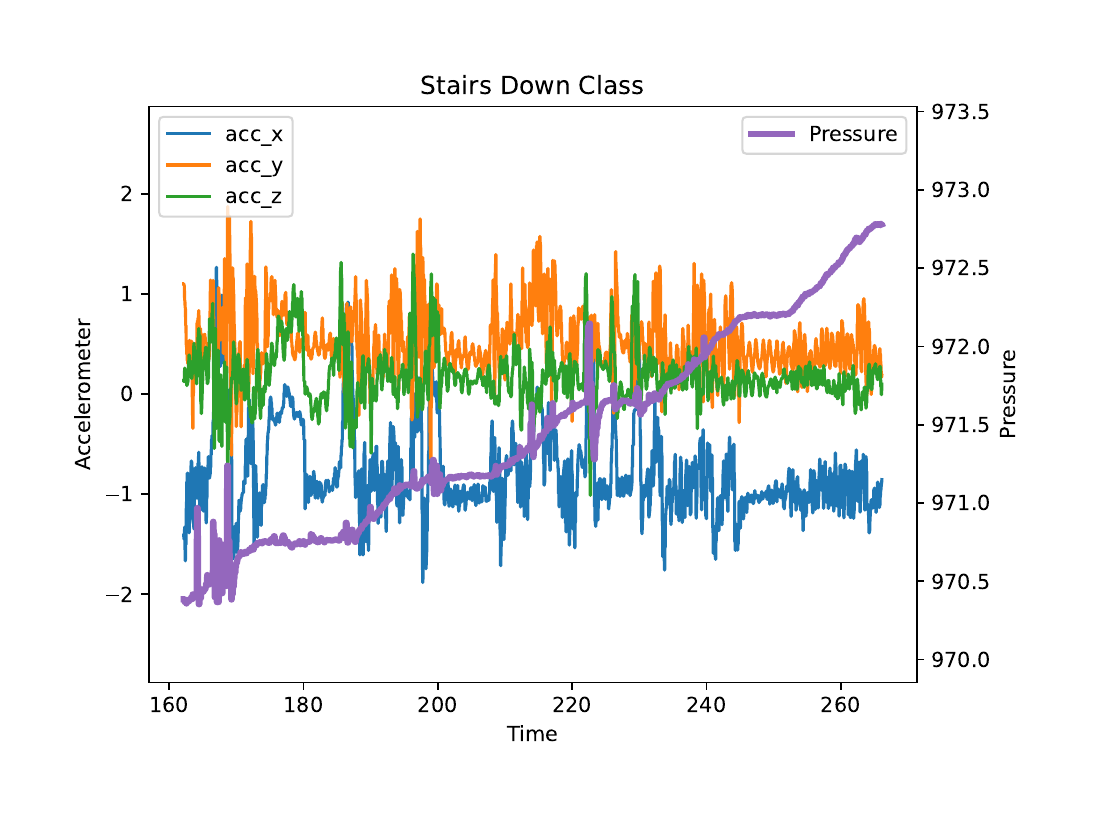}
    
    \caption{Example accelerometer and barometer data for the 'Stairs Down' activity class. The accelerometer shows a lot of activity, with similar walking patterns as in 'Stairs Up'. In direct comparison, the magnitude and frequency of the movement sensors seem higher. The barometer detects the steadily rising pressure, also interrupted by short plateaus caused by the landings.} 
    \label{fig:sensor_data-SD}
\end{figure}
\begin{figure}
    \centering
    \includegraphics[width=\linewidth,trim={1.2cm 1.2cm 1.2cm 1.2cm},clip]{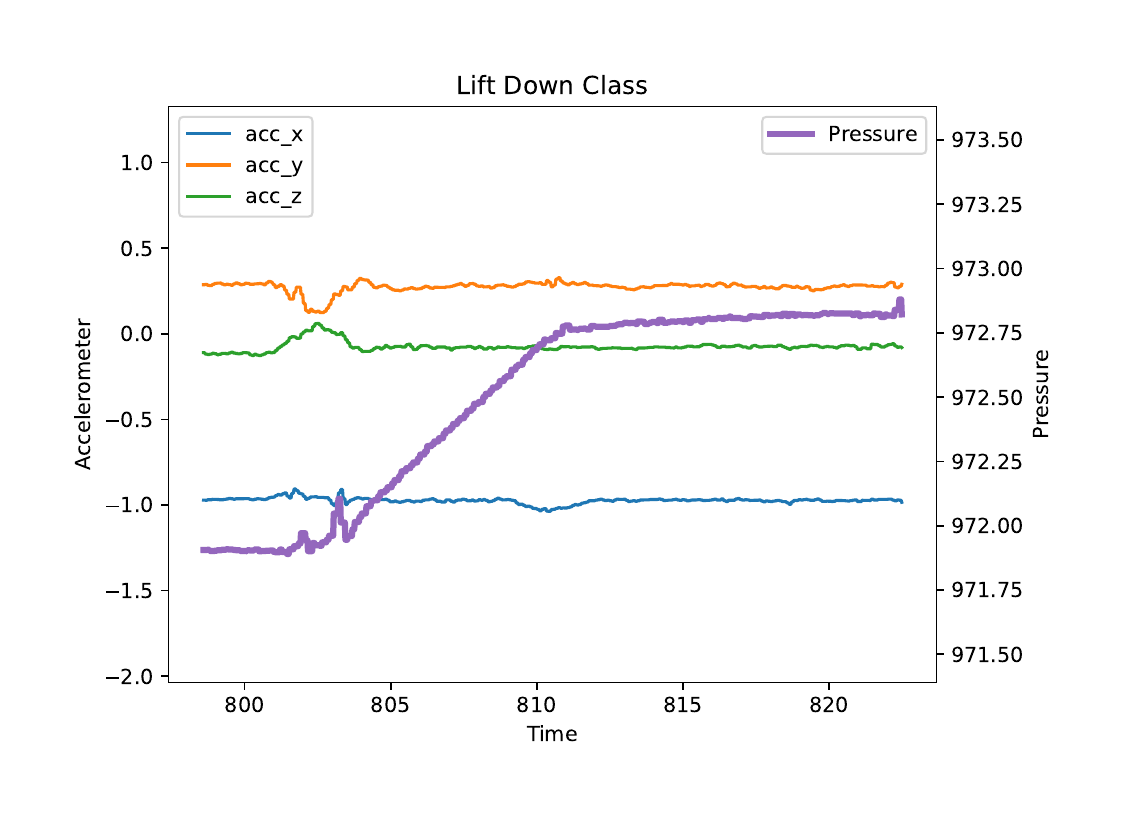}
    
    \caption{Example accelerometer and barometer data for the 'Lift Down' activity class. The specific participant for this instance was standing very still during the descent. We can see the pressure increasing linearly and smoothly.} 
    \label{fig:sensor_data-LD}
\end{figure}

\subsubsection{Annotation Data: }
The annotation data adds context to sensor data by designating specific time points as events or activities. This annotation can help segment or categorize sensor data based on various activities.
\begin{itemize}
    \item Elapsedtime: The elapsed time since the start of the recording.
    \item Comment: Annotations or labels provided at specific time points, such as "Lift down", "Stairs up", etc.
\end{itemize}

During the initial study of the sensor data, it was discovered that the annotation for "Lift" occurrences was completely dependent on the user entering the lift. However, it became clear that this approach did not accurately represent the user's experience because the lift may remain stationary for an extended amount of time before moving. This difference resulted in an erroneous portrayal of lift utilization in the data.

To address this issue and ensure the lift data's accuracy, it was decided to manually annotate certain data points within the sensor data that corresponded to lift occurrences. This manual annotation entailed carefully studying the sensor data and determining the exact instant when the lift began moving.

\section{Dataset Validation \& Prediction Algorithm}
When constructing a detection algorithm for evaluating sensor data, both the data separation method and the detection algorithm selection were carefully considered. To evaluate the collected dataset, a random forest classifier (from scikit-learn.ensemble libraries) has been chosen as the classification algorithm for this problem. Before feeding the raw data into the random forest classifier, we are pre-processing the data to extract features using sliding windows. We trained and evaluated the model on different window lengths of 4 and 8 seconds. For each of the windows, we are extracting 26 features from the data and the label is based on majority voting. During the window creation, windows are discarded if no label is assigned to at least 80\% of the total number of observations within the window. We also discard windows that are too short or are missing data, e.g. at the end of a recording.

We use leave-one-participant-out cross-validation to get a subject-independent performance evaluation.
In the prediction algorithm, the first step involves selecting a participant to serve as the test participant, while the remaining 19 participants are used for training the model. For these 19 participants, various features are computed as outlined in the data pre-processing stage. These features include statistical measures such as averages, minimums, maximums, variances, and standard deviations for accelerometer data (X, Y, Z), magnitude, and pressure, as well as additional metrics like range, slope, kurtosis, and skewness.

The class imbalance was addressed using the random oversampling technique, which duplicates randomly selected samples from the minority classes until all classes have the same number of samples. The \emph{imblearn} library has been used to oversample the data with 'not majority' as the sampling strategy. This step is meant to prevent the model from being biased towards the majority class and to ensure it performs well across all classes. For this proof-of-concept study, we did not explore the use of undersampling or a combination of oversampling and undersampling. A more sophisticated sampling technique like SMOTE could improve our model's performance further, but was out of the scope of this work.

To optimize the random forest classifier's performance, a grid search was conducted to identify the best hyperparameters, specifically the depth of the trees (max\_depth) and the number of trees in the forest (n\_estimators). Grid search involves testing various combinations of these hyperparameters to find the combination that yields the highest performance, in our case based on 10-fold cross-validation results. We used 'GridSearchCV' from the scikit-learn library. With the optimal hyperparameters identified, the random forest model was trained using the 19 participants in each LOSO split's training data.

Finally, for each left-out participant, the trained Random Forest model was then used to make predictions and evaluate its performance. The model's performance was evaluated by calculating the accuracy of its predictions, which measures the proportion of correctly classified instances out of the total number of instances. Additionally, we report the F1-score in three modes of multiclass-averaging (Micro, Macro, Weighted).

To measure the impact of the addition of the pressure sensor to the data, we performed a small ablation study, repeating the same training and validation procedure, but without the pressure data. This enabled us to test the hypothesis, that pressure data contains viable information to detect stairs and lift usage, but also to distinguish between these two, as well as their directions (up or down).

\section{Results \& Discussion}

We show the results of our machine learning experiments in Table \ref{tab:metrics_table_1}.
The results demonstrate significant performance improvements when the time-series data is pre-processed using an 8-second window for feature extraction compared to a 4-second window. The model's evaluation metrics, including accuracy and F1 scores, are notably higher with the 8-second window, reaching a macro F1 score of $0.86$ (8s) versus $0.76$ (4s). This substantial increase in accuracy highlights the importance of selecting an appropriate window size for pre-processing time-series data, as it can greatly enhance the classifier's ability to understand patterns and make accurate predictions. 

When comparing the performance of a Random Forest model trained solely on acceleration data (IMU data) to one that incorporates both acceleration and pressure data, the results clearly show that the inclusion of pressure data significantly enhances the model's ability to accurately distinguish between stair use and lift use. Without the barometric sensor, a macro F1-score of only 0.49 was reached. While acceleration data captures the dynamic movement patterns, pressure data can offer insights into altitude changes, such as those experienced when ascending or descending stairs. The altitude changes experienced when ascending or descending stairs are distinct from the relatively constant altitude changes during elevator use. 
For the evaluation metrics of each participant in both time-windows please refer to our GitHub repository.

\begin{table}[!h]
    \centering
    \begin{tabular}{|c|c|c|c|c|}
    \hline
        \textbf{Sensors} & \multicolumn{2}{c|}{\textbf{IMU \& Press.}} &  \multicolumn{2}{c|}{\textbf{IMU only}} \\ \hline
        \textbf{Time-window} & \textbf{8 s} & \textbf{4 s} & \textbf{8 s} & \textbf{4 s} \\ \hline
        \textbf{Accuracy} & 0.88 & 0.80 & 0.68 & 0.65\\ \hline
        \textbf{F1-Score (Micro-Avg)}& 0.88 & 0.80 & 0.68 & 0.65\\ \hline
        \textbf{F1-Score (Macro-Avg)}& 0.86 & 0.76 & 0.49 & 0.46\\ \hline
        \textbf{F1-Score (Weighted)} & 0.88 & 0.80  & 0.67 & 0.63\\ 
    \hline
    \end{tabular}
    \caption{The model evaluation metrics averaged over all 20 participants for 8-second and 4-second time-window. The results are shown with and without the use of the recorded pressure data. The columns on the left show results when IMU and the pressure sensor were used, while the columns on the right show the results without the pressure data.}
    \label{tab:metrics_table_1}
\end{table}

Evaluating F1 scores in micro, macro, and weighted modes offers a comprehensive understanding of the model's performance across different aspects of the data, especially when data is imbalanced. The macro F1 score is particularly informative in scenarios with class imbalance because it reveals the model's performance across each class individually. The weighted F1 score provides a balanced view that accounts for class imbalance while still reflecting the overall performance across all classes.
The table of complete metrics for each participant in 8-second and 4-second time windows along with the best parameter values for \emph{max\_depth} and \emph{n\_estimators} hyperparameters while fitting random forest classifier can be found in appendix B. \ref{tab:metrics_table_all1} and  C.\ref{tab:metrics_table_all2} respectively.

\begin{figure}[!h]
    \centering
    \includegraphics[width=.8\linewidth]{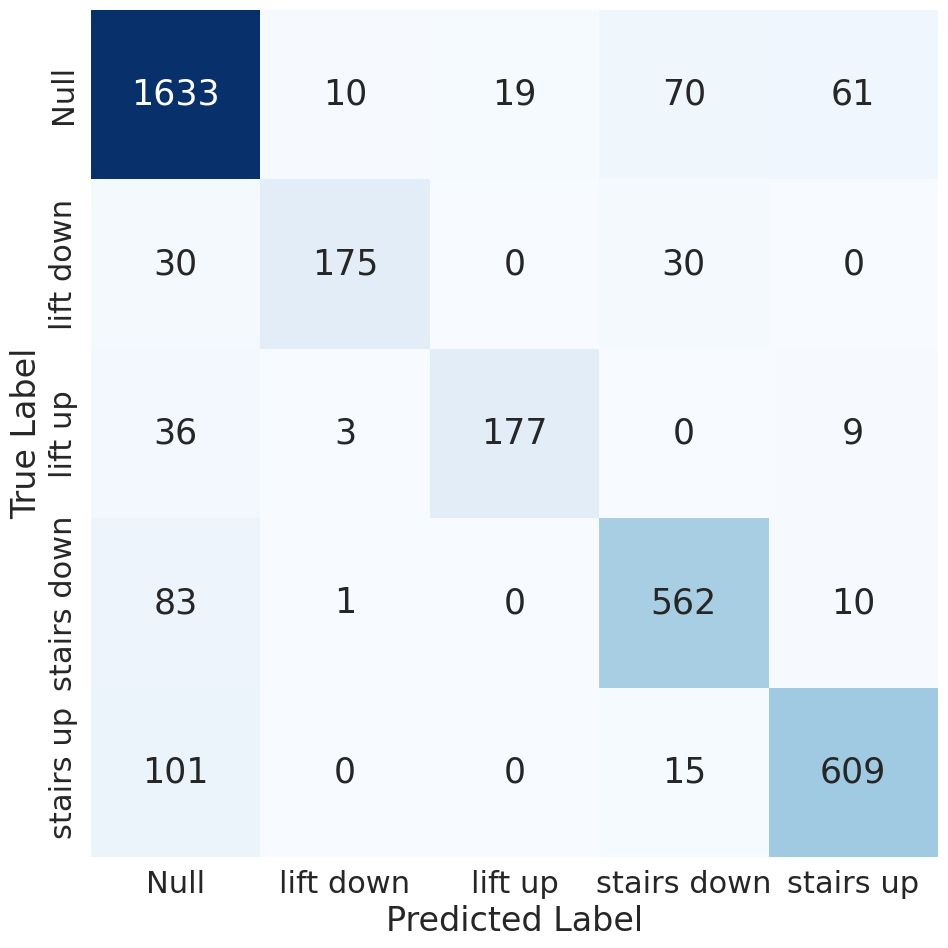}
    \caption{Confusion Matrix for 8-second windows. Most predictions are correct (diagonal of the matrix). We observe that most mistakes revolve around the null class and that 'stairs' and 'lift' can be separated reliably by the model.}
    \label{fig:confusion_matrix_1}
\end{figure}

Figures \ref{fig:confusion_matrix_1} and Appendix A Figure \ref{fig:confusion_matrix_2} present the confusion matrix of the model over an 8-second and a 4-second time window respectively. The confusion matrix offers a detailed breakdown of the model's performance by presenting the counts of true positives, true negatives, false positives, and false negatives for each class. In Figure \ref{fig:confusion_matrix_1}, we can observe that most prediction mistakes are related to the null class and that the classes related to taking stairs and taking the lift are rarely confounded.

In the feature importance analysis conducted across all 26 features for each of the 20 participants, it was observed that the feature "slope\_pressure" consistently exhibited the highest importance score. This finding underscores the significance of pressure values in effectively distinguishing between the five distinct classes. The steepness or gradient of pressure changes, as represented by the slope feature, appears to be particularly informative in differentiating between activities involving changes in elevation, such as ascending or descending stairs and using lifts. The feature importance scores of all 26 features are presented in the Appendix E as Table \ref{feature_scores2}.

Overall, the evaluation of the collected dataset with the explained methods shows that lift and stair usage can be distinguished and differentiated from the background class with a sufficiently high accuracy and F1 score, given the set of wearable sensors. 

\section{Conclusions}
This paper presents the possibility of differentiating movement into five different classes with acceleration and pressure readings from a smartwatch. The feature importance score analysis depicts the importance of pressure measurement in this particular classification problem. Barometric measuring is a sensing modality seldom used in HAR. This work serves as a proof of concept, highlighting the potential of utilizing pressure data in activity recognition. The findings indicate that pressure-related features, particularly the slope of pressure changes, are highly informative for detecting and classifying different movement patterns related to altitude changes. This insight could pave the way for more advanced and accurate activity recognition systems, which have applications in various applications such as health monitoring, fitness tracking, and smart home environments.

Furthermore, the effectiveness of different window sizes for pre-processing time-series data in the context of activity recognition using a random forest classifier is also investigated. Our findings highlight once more the critical role of window size selection in enhancing model performance. This underscores the importance of capturing more comprehensive temporal information to better discriminate between activity classes.

We believe the main contributions of our dataset and this paper's experiments are relevant to the Activity Recognition community:
\begin{itemize}
    \item A dataset dedicated specifically to taking lifts versus taking stairs detection was collected and annotated
    \item Our experiment serves as a proof of concept for the use of barometric air pressure data in Activity Recognition 
    \item We have investigated especially the window size in such time-series data pre-processing
    \item We analyzed Feature Importance scores to investigate the impact of both sensors and features
\end{itemize}
 
\section{Future work \& Enhancements}
 The dataset used in this study presents opportunities for further investigation, particularly in addressing imbalances among the five activity classes. Despite our efforts to collect data in an unbiased manner, the inherent variability in participants' activities has led to class imbalances. The lifts are faster than the stairs, which is one of the main reasons for the imbalance in data. In longer and real-world settings, the data is likely to be even more imbalanced. Future studies could explore strategies to mitigate this imbalance and to deal with it both during data collection and real-time inference.

Expanding the dataset size is another avenue for future research. Currently, data collection occurs within a single building, potentially limiting the generalizability of the model. By collecting data from diverse environments and settings, including outdoor and indoor scenarios, we can enhance the model's ability to generalize across different contexts. Additionally, increasing the dataset size by recruiting participants from various demographics and activity levels can enrich the diversity and representativeness of the dataset. With a larger dataset, implementing neural networks such as Long Short-Term Memory (LSTM) or transformer-based networks becomes feasible, allowing for the exploration of deep learning architectures capable of capturing even more complex temporal dependencies.

Furthermore, the definition of the Null class can be refined to incorporate a broader range of activities. While the null class primarily represents periods of inactivity, such as standing or sitting, future studies could consider including additional activities like cycling, jogging, or walking within this class. This expansion would provide a more comprehensive representation of daily activities, thereby improving the model's performance in real-world settings where activities may be more diverse and dynamic.

\begin{acks}
This project is funded by the Deutsche Forschungsgemeinschaft (DFG, German Research Foundation) – 425868829 and is part of Priority Program SPP2199 Scalable Interaction Paradigms for Pervasive Computing Environments.
\end{acks}

\bibliographystyle{ACM-Reference-Format}
\balance
\bibliography{main}

\newpage
\appendix
\section{Confusion Matrix for 4-seconds time-window}
\begin{figure}[h!]
    \centering
    \includegraphics[width=.8\linewidth]{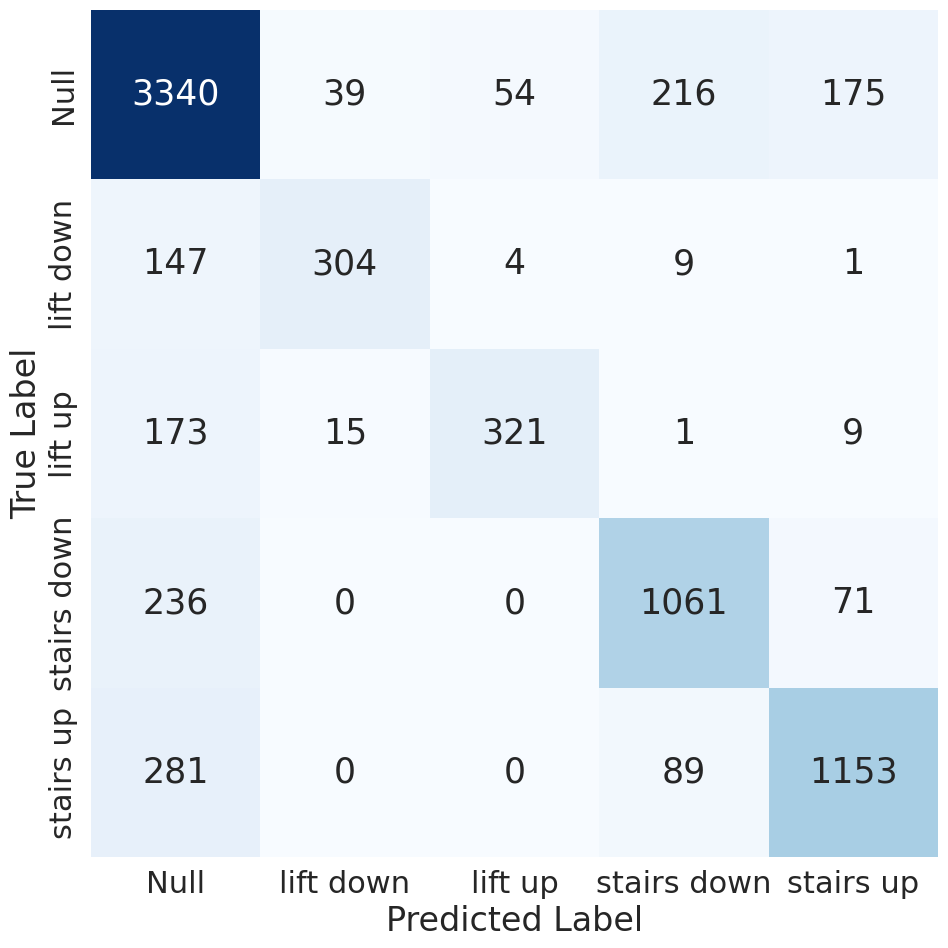}
    \caption{Confusion Matrix for 4-second windows.}
    \label{fig:confusion_matrix_2}
\end{figure}

\section{Evaluation Metrics for 20 participants in 8-seconds time-window} 
The following table presents evaluation metrics and also suitable hyper-parameter values for random forest found by GridSearchCV for each participant in 8-second windows.
    \begin{table}[H]
        \begin{tabular}{|c|c|c|c|c|c|}
        \hline
        No.& Acc.&macro F1&weighted F1&	Depth&	Est.s\\ \hline
        1& 0.84&	0.83&	0.84&	None&	200\\ \hline
        2& 0.85&	0.82&	0.85&	None&	250\\ \hline
        3& 0.90&	0.89&	0.90&	None&	250\\ \hline
        4& 0.82&	0.81&	0.83&	15&	275\\ \hline
        5& 0.96&	0.94&	0.96&	15&	250\\ \hline
        6& 0.87&	0.86&	0.87&	20&	200\\ \hline
        7& 0.81&	0.79&	0.81&   15&	225\\ \hline
        8& 0.89&	0.80&	0.88&	20&	200\\ \hline
        9& 0.90&	0.87&	0.90&	None&	275\\ \hline
        10& 0.91&	0.85&	0.90&	20&	250\\ \hline
        11& 0.92&	0.91&	0.92&	None&	300\\ \hline
        12& 0.90&	0.90&	0.90&	20&	200\\ \hline
        13& 0.87&	0.87&	0.86&	20&	300\\ \hline
        14& 0.89&	0.88&	0.89&	None&	250\\ \hline
        15& 0.87&	0.87&	0.87&	15&	350\\ \hline
        16& 0.88&	0.89&	0.88&	None&	300\\ \hline
        17& 0.86&	0.82&	0.86&	None&	300\\ \hline
        18& 0.88&	0.89&	0.88&	20&	300\\ \hline
        19& 0.86&	0.79&	0.86&	20&	275\\ \hline
        20& 0.85&	0.84&	0.85&	20&	225\\ \hline
    \end{tabular}
    \caption{The model evaluation metrics (Acc for Accuracy, Est.s for Estimators) for all 20 participants for 8-second time windows along with hyperparameter values.}
    \label{tab:metrics_table_all1}
\end{table}

\section{Evaluation Metrics for 20 participants in 4-seconds time-window} 
The following table presents evaluation metrics and also suitable hyper-parameter values for random forest found by GridSearchCV for each participant in 4-second windows.
    \begin{table}[!h]
        \centering
        \begin{tabular}{|c|c|c|c|c|c|}
        \hline
        No.& Acc.&macro F1&weighted F1&	Depth&	Est.s\\ \hline
        1&	0.77&		0.72&	0.77&	None&	350\\ \hline
        2&	0.81&	0.75&0.80&	None&	350\\ \hline
        3&	0.81&		0.76&0.81&	None&	250\\ \hline
        4&	0.77&	0.73&	0.78&	None&	225\\ \hline
        5&	0.78&		0.73&	0.78&	None&	200\\ \hline
        6&	0.79&		0.76&	0.79&	None&	200\\ \hline
        7&	0.75&		0.72&	0.76&	None&	250\\ \hline
        8&	0.78&		0.75& 0.77&	None&	350\\ \hline
        9&	0.85&		0.80&	0.84&	None&	250\\ \hline
        10&	0.84&		0.70&	0.83&	None&	350\\ \hline
        11&	0.87&		0.70&	0.83&	None&	350\\ \hline
        11&	0.87&			0.87&	0.87&	None&	325\\ \hline
        12&	0.80&		0.78&	0.80&	None&	325\\ \hline
        13&	0.80&		0.74&	0.79&	None&	250\\ \hline
        14&	0.80&		0.72&	0.80&	None&	325\\ \hline
        15&	0.83&		0.82&	0.83&	None&	275\\ \hline
        16&	0.81&		0.79&	0.81&	None&	300\\ \hline
        17&	0.79&		0.74&	0.79&	None&	225\\ \hline
        18&	0.77&	0.77&	0.77&	None&	275\\ \hline
        19&	0.83&		0.78&	0.83&	None&	275\\ \hline
        20&	0.77&	0.73	&0.76&	None&	200\\ \hline
    \end{tabular}
    \caption{The model evaluation metrics (Acc for Accuracy, Est.s for Estimators) for all 20 participants for 4-second time windows along with hyperparameter values.}
    \label{tab:metrics_table_all2}
\end{table}

\section{Total time duration for each class in minutes}
\begin{table}[!h]
    \centering
    \begin{tabular}{|c|c|}
    \hline
        \textbf{class} & \textbf{Minutes} \\ \hline
        Lift Down&	77.60\\ \hline
        Stairs Down	&59.20\\ \hline
        Stairs Up&	39.88\\ \hline
        Lift Up&	75.36\\ \hline
        Null&	272.98\\ \hline
    \end{tabular}
    \caption{Duration of data collected for each class in minutes}
    \label{duration_table}
\end{table}
 In total 152.95 minutes of data was collected for the lift classes and 99.08 minutes of data was collected for the stairs classes.

\newpage
\section{Feature-Scores for 26 features averaged over 20 participants in 8-seconds and 4-seconds time-window.}
    \begin{table}[!h]
        \centering
        \begin{tabular}{|c|l|l|}
        \hline
        \textbf{Feature} & \textbf{8-seconds} & \textbf{4-seconds}\\ \hline
        slope\_pressure& 0.313075423 & 0.24633301\\ \hline
        var\_magnitude& 0.084037245 & 0.079129592\\ \hline
        std\_magnitude&	0.082081804 & 0.072645593\\ \hline
        max\_magnitude&	0.045221187 & 0.051749068\\ \hline
        std\_pressure&	0.043930271 & 0.030623624\\ \hline
        var\_pressure& 0.043276678 & 0.031141125\\ \hline
        min\_magnitude& 0.036980905 & 0.032534554\\ \hline
        kurtosis\_pressure&	0.036261681 & 0.029589955\\ \hline
        min\_accX&	0.027751388 & 0.028021951\\ \hline
        range\_pressure& 0.02538081 & 0.025415817\\ \hline        
        std\_accX& 0.023843761 &	0.034228149\\ \hline
        var\_accX& 0.023416437 &	0.036206354\\ \hline
        std\_accZ&	0.020009935 & 0.02788877\\ \hline
        var\_accZ& 0.018040161 &	0.026410733\\ \hline
        avg\_accX& 0.016969519 &	0.02231106\\ \hline
        max\_accX&	0.016947775 & 0.022569319\\ \hline
        avg\_magnitude&	0.0169168 & 0.027153292\\ \hline       
        skew\_pressure&	0.015893485& 0.019383114\\ \hline
        min\_accZ&	0.015800581& 0.020498954\\ \hline
        max\_accY&	0.015392096& 0.020042188\\ \hline
        std\_accY&	0.014100437& 0.021778705\\ \hline
        var\_accY&	0.01405132& 0.021217197\\ \hline        
        avg\_accZ&	0.013379868 & 0.019155533\\ \hline
        min\_accY&	0.013372234 & 0.017988043\\ \hline
        avg\_accY&	0.012822567 & 0.017824296\\ \hline
        max\_accZ&	0.011045632 & 0.018160005\\ \hline
        \end{tabular}
        \caption{Feature scores for all features extracted over 8-seconds and 4-seconds window}
        \label{feature_scores2}
    \end{table}
\end{document}